\documentclass[conference, 9pt]{IEEEtran}
\IEEEoverridecommandlockouts
\usepackage{hhline}
\usepackage{colortbl}
\usepackage{centernot}
\usepackage{pbox}
\usepackage{amsmath}
\usepackage{amsfonts}
\usepackage{url}
\usepackage{bm}
\usepackage{comment}
\usepackage{graphicx}
\usepackage[hidelinks]{hyperref}
\usepackage{epstopdf}
\usepackage{multirow}
\usepackage{color}
\usepackage{tabu}
\usepackage{mdframed}
\usepackage{syntax}
\usepackage{fancyvrb}
\usepackage{subfig}
\usepackage{cleveref}
\crefname{figure}{fig.}{figures}
\Crefname{figure}{Fig.}{Figures}
\Crefname{table}{TABLE}{Tables}
\usepackage[ruled,linesnumbered,algo2e]{algorithm2e}
\usepackage[noend]{algpseudocode}
\usepackage{algorithmicx}
\usepackage{algpseudocode}
\usepackage{paralist}
\usepackage{stmaryrd}
\usepackage{tikz}
\usepackage[referable]{threeparttablex}
\usepackage{tabularx}
\usepackage{booktabs}
\usepackage{array}
\usepackage{fancyhdr}
\usepackage{float}
\usepackage{xspace}
\usepackage{pifont}
\usepackage{balance}
\usepackage{xcolor}% http://ctan.org/pkg/xcolor

\usepackage{tablefootnote}
\usepackage[normalem]{ulem}

\usepackage{tikz}
\usepackage{xcolor}

\usepackage{graphicx}
\usepackage{adjustbox}

\usepackage{float}

\newfloat{figtab}{htb}{fgtb}
\makeatletter
  \newcommand\figcaption{\def\@captype{figure}\caption}
  \newcommand\tabcaption{\def\@captype{table}\caption}
\makeatother

\definecolor{princetonorange}{RGB}{255,143,0}

\definecolor{lightgreen}{RGB}{198, 224, 183}
\definecolor{lightred}{RGB}{240, 205, 176}
\begin{document}

\title{NetEncoder: General Netlist Representation Learning via Self-Supervised Text-Attributed Graph}
\title{Towards Netlist Foundation Model: A Self-Supervised and Cross-Stage-Aware Netlist Encoder via Text-Attributed Graph}
\title{Towards Netlist Foundation Model: A Multimodal Cross-Stage-Aligned Netlist Encoder via Text-Attributed Graph}
% \font\myfont=cmr12 at 19pt
\title{Towards Netlist Foundation Model: \\ A Multimodal and Cross-Stage-Aligned Netlist Encoder \\ via Text-Attributed Graph}
\title{NetTAG: A Multimodal and Cross-Stage-Aligned \\ Netlist Foundation Model via Text-Attributed Graph}
\title{NetEncoder: A Multimodal RTL-and-Layout-Aligned \\ Netlist Foundation Model via Text-Attributed Graph}
\title{NetTAG: A Multimodal RTL-and-Layout-Aligned \\ \underline{Net}list Foundation Model via \underline{T}ext-\underline{A}ttributed \underline{G}raph}
% \title{RTLFusion: A Multimodal and Implementation-Aware \\ RTL Encoder for Multiple Pre-Synthesis PPA Prediction Tasks}

% \title{NetEncoder: A Multimodal and Cross-Stage-Aligned \\ Netlist Foundation Model via Text-Attributed Graph}

% \author{Wenji Fang, Wenkai Li, Shang Liu, Yao Lu, Hongce Zhang, Zhiyao Xie\textsuperscript{$\dagger$}}

% \affiliation{%
%  \vspace{.08in}
%  \institution{Hong Kong University of Science and Technology (HKUST)}
%  \vspace{.04in}
%  \institution{\{sliudx, yludf, wfang838, mengming.li\}@connect.ust.hk, \ \ eezhiyao@ust.hk}
%  \country{}
%  \vspace{.03in}
% }

\author{
    \IEEEauthorblockN{
        Wenji Fang$^1$,
        Wenkai Li$^1$,
        Shang Liu$^1$,
        Yao Lu$^1$,
        Hongce Zhang$^2$,
        Zhiyao Xie$^1$$\textsuperscript{*}$
    }
    \IEEEauthorblockA{ $^1$Hong Kong University of Science and Technology}
    \IEEEauthorblockA{ $^2$Hong Kong University of Science and Technology (Guangzhou)}
    {$\textsuperscript{*}$Corresponding Author\vspace{-5pt}}
}

% \author{
% \IEEEauthorblockN{Wenji Fang, Wenkai Li, Shang Liu, Yao Lu, Hongce Zhang, Zhiyao Xie}
% % \IEEEauthorblockA{\textit{dept. name of organization (of Aff.)} \\
% % \textit{name of organization (of Aff.)}\\
% % City, Country \\
% % email address or ORCID}
% % \and
% % \IEEEauthorblockN{2\textsuperscript{nd} Given Name Surname}
% % \IEEEauthorblockA{\textit{dept. name of organization (of Aff.)} \\
% % \textit{name of organization (of Aff.)}\\
% % City, Country \\
% % email address or ORCID}
% }
% \affil[]{\fontsize{10}{10}\selectfont $^1$Hong Kong University of Science and Technology, \\$^2$Hong Kong University of Science and Technology, $^3$Duke University\vspace{-6pt}}

% \author{ \fontsize{11}{11}\selectfont Wenji Fang$^{1,2}$, Wenkai Li, Shang Liu, Yao Lu, Hongce Zhang$^{1,2}$, Zhiyao Xie$^2$\textsuperscript{*}\vspace{-5pt}}

% \affil[]{\fontsize{10}{10}\selectfont $^1$Hong Kong University of Science and Technology (Guangzhou), \\$^2$Hong Kong University of Science and Technology, $^3$Duke University\vspace{-6pt}}

% \affil[]{$\textsuperscript{*}$Corresponding Author: eezhiyao@ust.hk\vspace{-12pt}}

\maketitle

\begin{abstract}
Circuit representation learning has shown promise in advancing Electronic Design Automation (EDA) by capturing structural and functional circuit properties for various tasks. Existing pre-trained solutions rely on graph learning with complex functional supervision, such as truth table simulation. However, they only handle simple and-inverter graphs (AIGs), struggling to fully encode other complex gate functionalities. While large language models (LLMs) excel at functional understanding, they lack the structural awareness for flattened netlists.
To advance netlist representation learning, we present NetTAG, a netlist foundation model that fuses gate semantics with graph structure, handling diverse gate types and supporting a variety of functional and physical tasks. Moving beyond existing graph-only methods, NetTAG formulates netlists as text-attributed graphs, with gates annotated by symbolic logic expressions and physical characteristics as text attributes. Its multimodal architecture combines an LLM-based text encoder for gate semantics and a graph transformer for global structure. Pre-trained with gate and graph self-supervised objectives and aligned with RTL and layout stages, NetTAG captures comprehensive circuit intrinsics. Experimental results show that NetTAG consistently outperforms each task-specific method on four largely different functional and physical tasks and surpasses state-of-the-art AIG encoders, demonstrating its versatility.\looseness=-1
\end{abstract}

% \begin{IEEEkeywords}
% component, formatting, style, styling, insert.
% \end{IEEEkeywords}

\maketitle

\section{Introduction}

Machine learning (ML) techniques have demonstrated remarkable achievements in electronic design automation (EDA). 
Existing ML methods are mostly tailored to specific tasks such as predicting physical metrics (e.g, timing~\cite{fang2024annotating, wang2023restructure, guo2022timing}, area~\cite{fang2023masterrtl, xu2023fast, sengupta2022good}, power~\cite{du2024powpredict, xie2021apollo, zhou2019primal}, and routability~\cite{xie2018routenet, liu2021global, zheng2023lay}) or the reasoning of circuit functionalities~\cite{wu2023gamora, alrahis2021gnn, chowdhury2021reignn, he2021graph}. These methods are typically developed by supervised training, requiring extensive label collection and model customization for each task.
Despite obvious effectiveness, they are time-consuming to develop and lack generalizability, often capturing only task-specific patterns rather than a generalizable understanding of circuits.\looseness=-1

Recent circuit representation learning methods (i.e., encoders) generate informative embeddings for circuits and are able to support a range of downstream tasks~\cite{chen2024large, fang2025cfm}. 
Existing methods mainly focus on VLSI circuit netlists, and some works also explore RTL circuits~\cite{fang2025circuitfusion}, analog circuits~\cite{zhu2022tag}, and FPGAs~\cite{sohrabizadeh2023robust}. We provide a detailed summary of the netlist encoders in~\Cref{tbl:works}.
All these netlist encoders capture circuit functionality through graph structure leveraging graph learning models, such as Graph Neural Networks (GNNs) or Graph Transformers (GTs). They either directly infer functionality with graph topology~\cite{luo2024hnn, deng2024less, yang2022versatile}, or employ functional information to pre-train the models~\cite{li2022deepgate, shi2023deepgate2, shi2024deepgate3, wang2022functionality, wang2024fgnn2, khan2024deepseq, fang2025circuitencoder}.\looseness=-1
% They employ either a supervised strategy~\cite{luo2024hnn, deng2024less, yang2022versatile} or a pretrain-finetune paradigm~\cite{li2022deepgate, shi2023deepgate2, shi2024deepgate3, wang2022functionality, wang2024fgnn2, khan2024deepseq} to capture circuit information.

% They either employ supervised methods or use the pretrain-finetune paradigm methods leveraging self-supervised techniques or circuit functional supervisions to pre-train the graph learning models such as Graph Neural Networks (GNNs) or Graph Transformers (GTs). 

However, we argue that existing graph-only netlist encoders lack sufficient expressiveness to capture complex circuits. \textbf{These methods rely on graph models that inherently prioritize \textit{structural} information over \textit{semantics} (i.e., functionality).} Their limitations include: \textit{(1) Limited to and-inverter graph (AIG).} Most methods target only simple AIG-based netlists, representing a narrow subset of standard cell libraries, and cannot handle post-mapping netlists with diverse gate types.
\textit{(2) Dependence on complex functional supervision.} Representative methods like the DeepGate family~\cite{li2022deepgate, shi2023deepgate2, shi2024deepgate3, khan2024deepseq} and FGNN~\cite{wang2022functionality, wang2024fgnn2} incorporate truth table simulation to capture AIG functionality. However, truth tables suffer from exponential state expansion when applied to post-synthesis netlists with multi-input complex gates (e.g., full adders, multiplexers), limiting their applicability.
\textit{(3) Lack of physical information.} None of these methods consider physical characteristics, focusing solely on logic functionality.
Consequently, these encoders are limited to AIG-based tasks (e.g., equivalence checking, SAT solving) and struggle with tasks for post-mapping netlists~\cite{wang2023restructure, guo2022timing, du2024powpredict, xie2018routenet, liu2021global, zheng2023lay, alrahis2021gnn, chowdhury2021reignn}, especially those evaluating physical design quality.\looseness=-1

% These limitations restrict such AIG encoders to a narrow range of functional tasks  involving only and-inverter gates. Consequently, their applicability to numerous critical post-mapping netlist tasks~\cite{wang2023restructure, guo2022timing, du2024powpredict, xie2018routenet, liu2021global, zheng2023lay, alrahis2021gnn, chowdhury2021reignn} is significantly limited, particularly those evaluating physical design quality for sequential circuits.

% Despite the achievements of circuit representation learning, a critical challenge still remains: \textit{How can we develop a more generalizable netlist encoder that captures circuit intrinsic properties, encoding diverse gate types to support various netlist-stage tasks?}

Recently, large language models (LLMs) have demonstrated remarkable expressiveness in circuit-related generative tasks~\cite{pei2024betterv, fang2024assertllm, wu2024chateda}. \textbf{However, LLMs inherently capture circuit textual \textit{semantics} rather than \textit{structure}.} Limited work has explored LLMs for general-purpose circuit representation learning, with key limitations summarized as follows:
\textit{(1) Struggle with netlists.} Although LLMs can interpret circuit functionality from RTL code, the gate-level netlists are more flattened and lack informative context, making functional understanding more challenging.
\textit{(2) Lack of structural encoding.} LLMs struggle to capture the circuit structures, limiting their utility for netlist representation learning.

\begin{figure}[!t]
  \centering
  \vspace{-.1in}
  \includegraphics[width=1\linewidth]{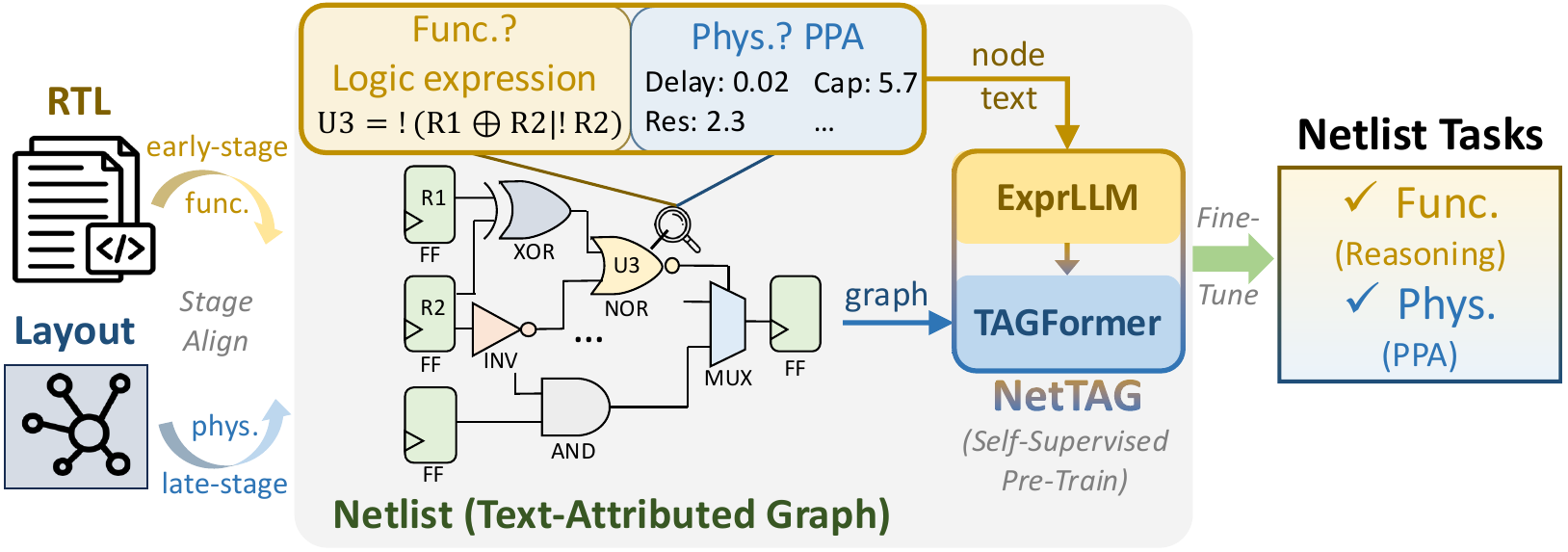}
  \vspace{-.2in}
  \caption{Overvew of NetTAG. Netlists are formulated as text-attributed graphs, with functional and physical text attributes extracted for each gate. Within NetTAG, gate attributes are initially encoded by ExprLLM, then refined with global netlist graph structures using TAGFormer. NetTAG is pre-trained with self-supervised objectives and aligned with RTL and layout embeddings, enabling versatile support for both functional and physical tasks after fine-tuning.\looseness=-1}
  \label{fig:motivation}
  \vspace{-.2in}
\end{figure}

% Most works target the netlist stage, learning the embedding of each netlist gate. However, existing works are not general enough to handle realistic netlists. All these works can only deal with And-Inverter Graphs (AIGs), overlooking other types of gates, which are crucial for the physical design quality prediction tasks. Additionally, most work only targets the combinational components 

In this work, we present \textbf{\texttt{NetTAG}}, a foundation model\footnote{The code and pre-trained NetTAG model are available at https://github.com/hkust-zhiyao/NetTAG. The pre-trained model enables users to easily generate and fine-tune embeddings for their own netlist tasks.} for netlists that captures functional and physical properties across diverse gate types. 
Unlike previous circuit encoders and LLM solutions focusing on single circuit modality, NetTAG fuses gate text \textit{semantics} with global graph \textit{structures} to achieve \textit{functional} and \textit{physical} understanding.
Serving as a foundation model, the pre-trained NetTAG generates versatile embeddings for logic gates, register cones, and full circuits. 
The embeddings can be easily fine-tuned for various downstream tasks, supporting largely different netlist-stage functional and physical tasks across these circuit granularities. 
As shown in~\Cref{fig:motivation}, we propose the following innovative strategies in NetTAG:\looseness=-1

\begin{table*}[!t]
\centering
\vspace{-.3in}
\caption{Comparision of state-of-the-art netlist representation learning methods.}
\vspace{-.05in}
\resizebox{0.8\textwidth}{!}{

\begin{tabular}{c||cc|cccc|cc} \toprule
\multirow{2}{*}{\textbf{Method}} &
  \multicolumn{2}{c|}{\textbf{Target Circuit}} &
  \multicolumn{4}{c|}{\textbf{Encoding Methodology}} &
  \multicolumn{2}{c}{\textbf{Downstream Tasks}} \\ \cline{2-9}
 &
  Cell Type &
  Circuit Type &
  Modality &
  ML Model &
  Pre-Train Target &
  Cross-Stage Align &
  Target &
  Type \\ \hline \hline
DeepGate1/2~\cite{li2022deepgate, shi2023deepgate2} &
  AIG &
  Comb. &
  Graph &
  GNN &
  Gate &
  N/A &
  Gate &
  Func. \\ \hline
DeepGate3~\cite{shi2024deepgate3} &
  AIG &
  Comb. &
  Graph &
  GT &
 Gate\&Circuit &
  N/A &
  Gate &
  Func. \\ \hline
FGNN~\cite{wang2022functionality,   wang2024fgnn2} &
  AIG &
  Comb. &
  Graph &
  GNN &
  Circuit &
  N/A &
  Gate\&Circuit &
  Func. \\ 
  \hline
HOGA~\cite{deng2024less} &
  AIG &
  Comb.\&Seq. &
  Graph &
  GNN &
  N/A &
  N/A &
  Gate\&Circuit &
  Func.\&Phys. \\ 
  \midrule
\textbf{\cellcolor[HTML]{FFF2CC}\textbf{\begin{tabular}[c]{@{}c@{}}NetTAG\\      (Ours)\end{tabular}}}&
  \cellcolor[HTML]{FFF2CC}\textbf{\begin{tabular}[c]{@{}c@{}}Any\\      Gate\end{tabular}}&
  \cellcolor[HTML]{FFF2CC}\textbf{\begin{tabular}[c]{@{}c@{}}Comb.\&\\      Seq.\end{tabular}} &
  \cellcolor[HTML]{FFF2CC}\textbf{\begin{tabular}[c]{@{}c@{}}Text\&\\      Graph\end{tabular}} &
  \cellcolor[HTML]{FFF2CC}\textbf{\begin{tabular}[c]{@{}c@{}}LLM\&\\      GT\end{tabular}} &
  \cellcolor[HTML]{FFF2CC}\textbf{\begin{tabular}[c]{@{}c@{}}Gate\&\\      Reg. Cone\end{tabular}} &
  \cellcolor[HTML]{FFF2CC}\textbf{\begin{tabular}[c]{@{}c@{}}RTL\&\\      Layout\end{tabular}} &
  \cellcolor[HTML]{FFF2CC}\textbf{\begin{tabular}[c]{@{}c@{}}Gate\&\\      Reg. Cone\&\\      Circuit\end{tabular}} &
  \cellcolor[HTML]{FFF2CC}\textbf{\begin{tabular}[c]{@{}c@{}}Func.\&\\      Phys.\end{tabular}} \\ \bottomrule
\end{tabular}

}

\label{tbl:works}
\vspace{-.2in}
\end{table*}

\begin{itemize}
    \item \textbf{Preprocess: formulating netlists as text-attributed graphs.} This paper is the first to represent netlists in text-attributed graphs (TAGs) format. 
    In the netlist graph, we annotate each gate with functional symbolic logic expressions and physical characteristics as node-level text attributes. 
    Our TAG format combines gate textual attributes with graph topology, moving beyond existing graph-only circuit representation learning. 
    \item \textbf{Multimodal architecture: fusing semantic and structure.} Leveraging TAGs, NetTAG introduces an innovative multimodal architecture that combines an LLM-based text encoder (i.e., ExprLLM) with a graph transformer (i.e., TAGFormer). ExprLLM first encodes \textit{fine-grained} gate text attributes into semantic-rich node embeddings, then TAGFormer refines them with \textit{overall} graph structure, generating final individual gate and overall graph embeddings.\looseness=-1
    \item \textbf{Aligned pre-training: self-supervised and cross-stage-aware.} We introduce four circuit-specific self-supervised objectives to pre-train NetTAG across circuit granularities, enhancing Boolean logic understanding and fusing gate semantics with the global graph structure.
    Moreover, we align netlist embeddings with those from RTL and layout stages to strengthen cross-stage functional and physical awareness.
\end{itemize}

We evaluate NetTAG on four largely different EDA tasks, covering both functional reasoning and physical design quality predictions at different circuit granularities. NetTAG consistently outperforms all state-of-the-art (SOTA) task-specific models, achieving 14\% and 13\% higher accuracy in logic gate and register function identification, respectively, and reducing errors by 2\% for register slack, and 7\% for both circuit area and power predictions.
NetTAG demonstrates strong generalization across these representative tasks.
Finally, we evaluate the scalability of NetTAG by scaling up model size and dataset volume. The results indicate a huge potential for improvement by scaling up either model or circuit datasets in future work.\looseness=-1

\section{Methodology}\label{sec:method}

\subsection{Overview}
\begin{figure}[!t]
  \centering
  \includegraphics[width=1\linewidth]{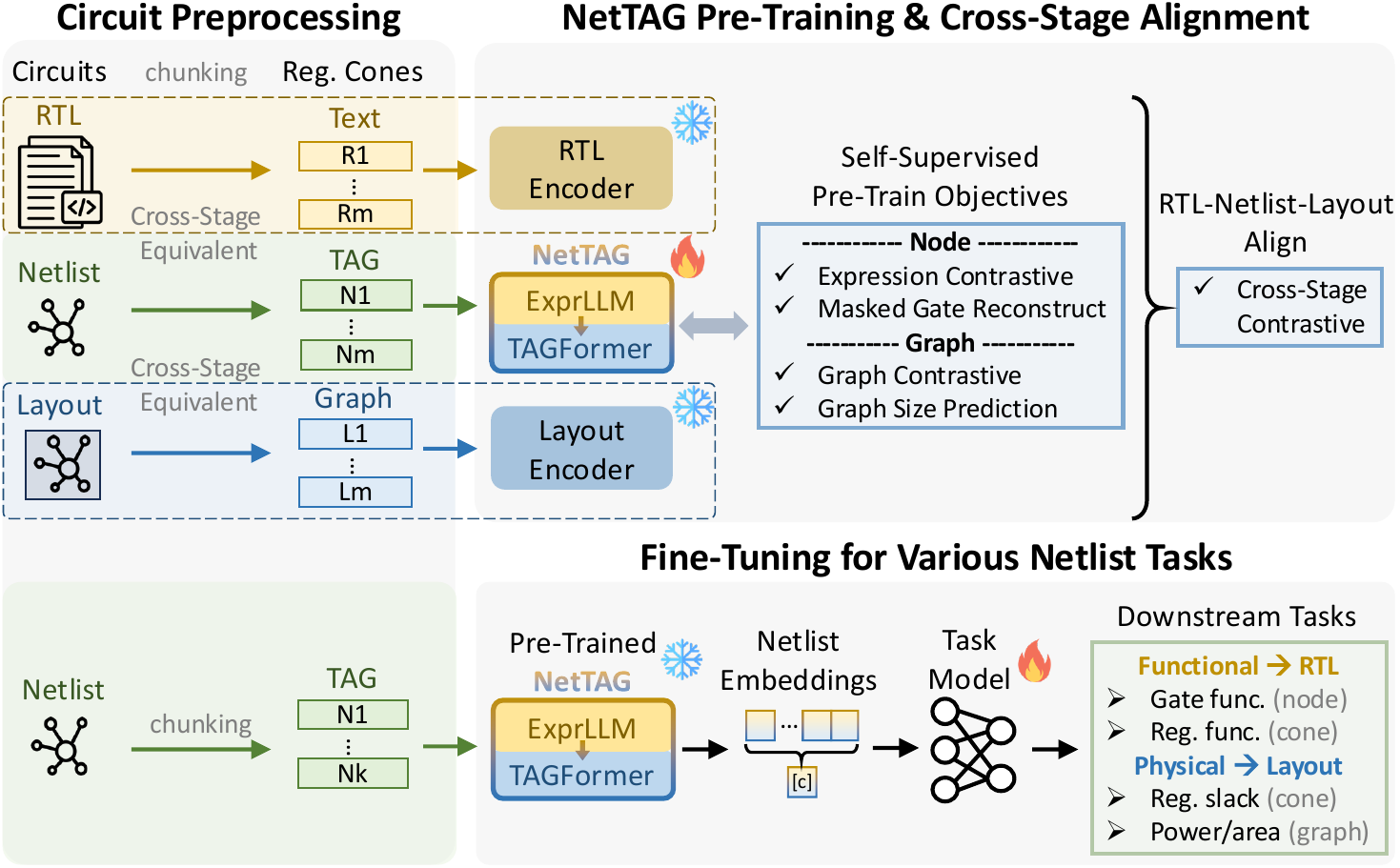}
  \vspace{-.2in}
  \caption{NetTAG Workflow. Sequential netlists are chunked into combinational register cones and converted into TAGs. During pre-training, NetTAG is trained with node-level and graph-level self-supervised objectives, and it is aligned with RTL and layout embeddings. 
  The pre-trained NetTAG then generates netlist embeddings, which are fine-tuned with netlist-stage task labels.}
  \label{fig:flow}
  \vspace{-.2in}
\end{figure}

\Cref{fig:flow} outlines the NetTAG workflow. First, netlist data is converted into TAG format, detailed in~\Cref{sec:preproc}. NetTAG’s multimodal architecture, which combines fine-grained gate text and overall circuit graph refinement, is illustrated in~\Cref{sec:model}. Step 1 enhances Boolean logic comprehension for ExprLLM (\Cref{sec:pretrain1}), and Step 2 fuses gate semantics with circuit structure in TAGFormer, incorporating cross-stage alignment to strengthen both functional and physical understanding (\Cref{sec:pretrain2}). Finally, \Cref{sec:finetune} demonstrates the fine-tuning strategy that leverages the pre-trained NetTAG for various downstream tasks.\looseness=-1

% This section introduces our NetTAG framework in detail. Given a gate-level netlist $\mathcal{N}$, we represent it as a text-attributed graph $\mathcal{G}_\mathcal{N} = \{\mathcal{T}, \mathcal{E}\}$, where each gate in the netlist is a node in the graph. Here, $\mathcal{T} = \{t_1, \dots, t_n\}$ is a set of gate text attributes, where $n$ representing the total gate number. Each attribute $t_i$ contains functional and physical properties of its gate, consisting of a sequence of tokens $t_i = \{w_1, \dots, w_m\}$, where $m$ is the total token count. $\mathcal{E} = \{e_{ij}\}$ is a set of edges connecting the gates.

% Our goal is to design a netlist encoder that generates netlist embeddings $N$ that capture both gate textual semantics in $\mathcal{T}$ and graph connectivity in $\mathcal{E}$, as formulated below: 
% \begin{equation}
% N = \text{NetTAG}({\{\mathcal{T}, \mathcal{E}\}}).
% \end{equation}

% The framework of NetTAG is demonstrated in~\Cref{fig:flow}. The RTL-netlist-layout equivalent circuit data is first processed into corresponding modalities: the target netlist is transformed into TAG, and cross-stage equivalent RTL code is directly handled as text while its layout is converted into graph with physical information (see~\Cref{sec:preproc}). Then the netlist TAG is encoded by our NetTAG, with node text encoded by ExprLLM into initial node embeddings and TAGFormer further enhances the embeddings by capturing the structural information. The RTL and layout are also encoded by corresponding encoders. 

\subsection{Circuit Data Preprocessing} \label{sec:preproc}

\begin{figure*}[!t]
  \centering
  \vspace{-.3in}
  \includegraphics[width=1\linewidth]{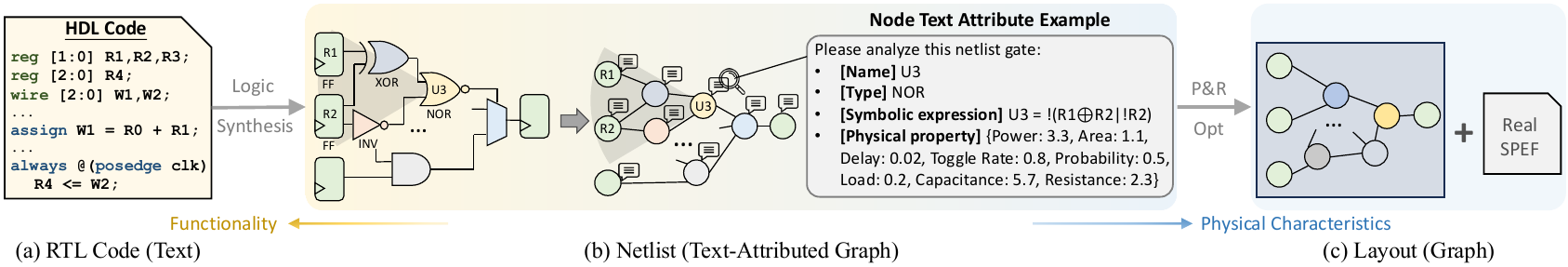}
  \vspace{-.2in}
  \caption{Circuit data design stages and modalities. 
  RTL code is processed directly as text. 
  Netlists are represented as TAGs, with node attributes including gate name, type, symbolic expression, and physical property. Symbolic expressions are derived from each gate’s k-hop input cone.
  Layout data is converted into graphs and annotated with physical information extracted from the SPEF file.}
  \label{fig:align}
  \vspace{-.2in}
\end{figure*}

Conventional graph methods like GNNs and GTs mainly capture structural information but fall short when the semantic context is also needed. TAG representation learning bridges this gap by encoding nodes as text and leveraging graph connections~\cite{he2024harnessing,liu2024one}, effectively improving node classification performance.
\textit{Intuitively, netlists, with their graph structure and detailed gate logic functions, are ideally suited for formulation as TAGs.}

\textbf{Formulating netlist as TAG.} 
Given a gate-level netlist $\mathcal{N}$, we represent it as a text-attributed graph $\mathcal{G}_\mathcal{N} = \{\mathcal{T}, \mathcal{E}\}$, where each gate in the netlist is a node in the graph. Here, $\mathcal{T} = \{t_1, \dots, t_m\}$ is a set of gate text attributes, where $m$ representing the total gate count. Each attribute $t_i$ contains functional and physical properties of the $i^{th}$ gate, as shown in~\Cref{fig:align}(b), and consists of a sequence of tokens $t_i$. $\mathcal{E}$ is the set of edges connecting the gates.

Specifically, for functionality, we extract the symbolic logic expression for each gate based on its k-hop fan-in cone. These expressions are constructed using a formal verification toolkit~\cite{gario2015pysmt}, providing formal definitions for gate functionality.
An example is demonstrated in~\Cref{fig:align}(b), the 2-hop expression for this NOR gate with the symbolic name ``U3'' is derived from the Boolean functions of nodes within its 2-hop fan-in cone. The expression for ``U3'' is represented as:
``$\text{U3} = !((\text{R1} \oplus \text{R2}) | ! \text{R2})$''. 
Each expression includes symbolic names for input gates and the Boolean operations.
We summarize the key advantages of exploiting the symbolic expression for functional encoding:\looseness=-1 
\begin{enumerate}
    \item \textbf{Versatile gate support:} Our symbolic expressions use Boolean formulas to represent any gate type, including complex gates like AOI and full adders. This enables NetTAG to support diverse gate types in netlists, beyond the limitations of AIG formats.\looseness=-1 
    \item \textbf{Direct functional encoding:} In contrast to pre-trained circuit encoders that rely on graph-based learning with truth table supervision~\cite{wang2024fgnn2, shi2024deepgate3}, symbolic expressions enable straightforward static analysis, covering all input conditions without exponential growth problems by exhaustive truth table simulation.
    \item \textbf{Structure independence:} Unlike structure-based methods that infer functionality with graph topology~\cite{deng2024less, yang2022versatile}, our approach derives gate expressions from Boolean formulas. It easily recognizes the same functionality across varying structures and distinguishes different functions in similar topologies.
    \item \textbf{Compatible with language models:} Expressions can be seamlessly processed by language models, unleashing LLMs' reasoning capabilities to enhance circuit functional understanding. This functional expression also enables self-supervised pre-training for LLMs, such as expression contrastive learning, to improve LLMs’ logic understanding from expression data itself.
\end{enumerate}

For the physical characteristics, we annotate each gate with information extracted from the standard cell library, including characteristics such as power, area, delay, toggle rate, probability, load, capacitance, and resistance. Both functional and physical attributes are combined to form the text attribute for each gate.

\textbf{Processing RTL and layout data.} In addition to the target netlists, we utilize corresponding RTL and layout data to enhance NetTAG’s cross-stage awareness. As shown in~\Cref{fig:align}(a) and (c), RTL code is processed directly as text containing functional semantics, while layout data is represented as connectivity graphs annotated with physical characteristics. Specifically, nodes in the layout graphs are annotated with capacitance, resistance, and delay values extracted from the Standard Parasitic Extraction Format (SPEF) file.

\textbf{Chunking sequential circuit into register cones.} 
To handle large-scale sequential circuits, we propose chunking the original circuit into register cones. For each register, we backtrace through all driving logic up to other registers, creating a subcircuit cone that captures the complete state transition of the register, including update relationships and timing paths. This chunking process significantly reduces circuit size for our NetTAG, enhancing scalability. We apply this method consistently across netlist, RTL, and layout data, ensuring that cross-stage cones remain functionally equivalent\footnote{Most registers maintain unchanged throughout design stages.} for subsequent alignment.

\subsection{NetTAG Model Architecture} \label{sec:model}
\begin{figure*}[!t]
  \centering
  \vspace{-.3in}
  \includegraphics[width=0.95\linewidth]{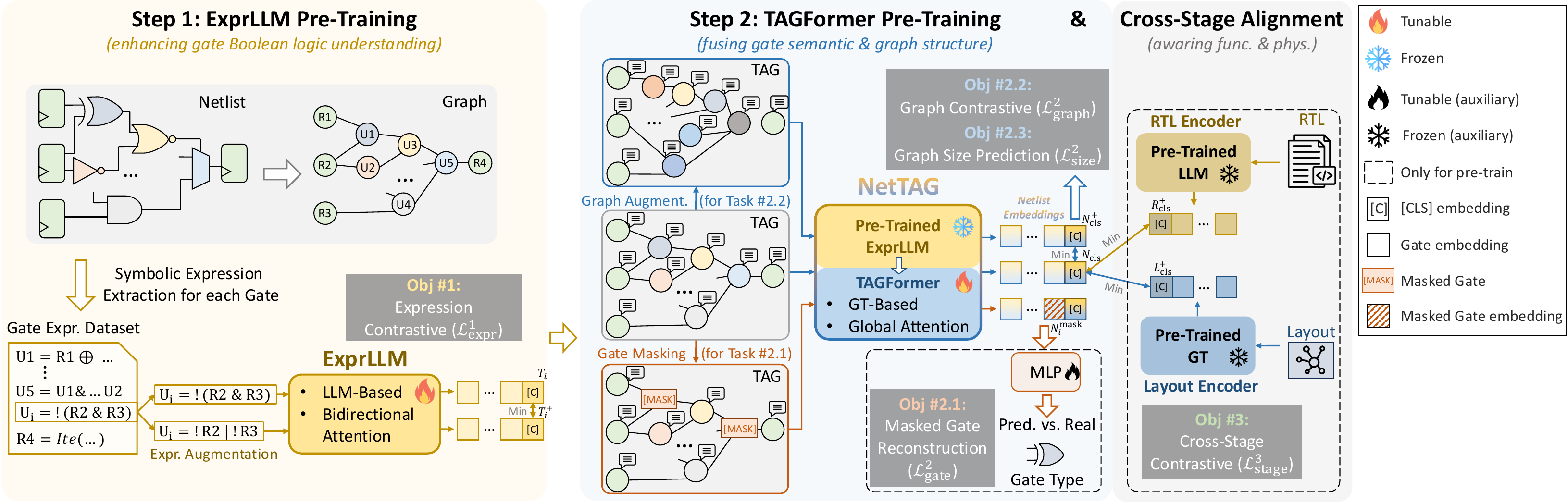}
  \vspace{-.05in}
  \caption{NetTAG architecture and pre-training workflow. 
  In step 1, we collect a dataset of gate symbolic expressions and pre-train ExprLLM to enhance its understanding of Boolean formulas. In step 2, ExprLLM is frozen, and we pre-train TAGFormer to fuse the gate semantics with the global graph structure. NetTAG embeddings are cross-stage aligned with those from pre-trained auxiliary RTL and layout encoders.}
  \label{fig:pretrain}
  \vspace{-.2in}
\end{figure*}

\Cref{fig:pretrain} illustrates the architecture of NetTAG, which comprises two main components: ExprLLM for gate-level text encoding and TAGFormer for further circuit-level graph encoding. Additionally, two auxiliary pre-trained encoders are employed to generate embeddings for the RTL code and layout graph, facilitating cross-stage alignment. We detail our NetTAG architecture below.

\textbf{ExprLLM of NetTAG} is initialized using an LLM-based text embedding model to encode gate text attributes as initial node embeddings. 
Following~\cite{behnamghader2024llm2vec}, we adapt a decoder-only LLM into a general-purpose text encoder by converting causal attention to bidirectional attention, which has shown superior encoding performance over encoder-only models (e.g., BERT)~\cite{muennighoff2023mteb}.
ExprLLM takes the gate text attributes $t_i$ as input, converting them into the text embeddings $T_i$:\looseness=-1
\vspace{-.05in}
\begin{equation}
T_i = \text{ExprLLM}(t_i).
\end{equation}

\textbf{TAGFormer of NetTAG} refines node embeddings from ExprLLM using a graph transformer model, which captures netlist graph structure via global attention.
We include a \texttt{[CLS]} node connected to all other nodes in the graph, serving as the graph-level embedding.
The initial embedding for each gate $n_i$ is created by concatenating its text embedding $T_i$ from ExprLLM with its physical characteristics vector $x_{\text{phys}i}$. TAGFormer then processes node embeddings $\{n_1, \dots, n_m\}$ with their connectivity $\mathcal{E}$ into a sequence of netlist gate embeddings $\{N_1, \dots, N_m, N_{\text{cls}}\}$, where $N_{\text{cls}}$ represents the entire graph embedding:\looseness=-1
\vspace{-.05in}
\begin{equation}
\begin{split}
n_i &= (T_i, x_{\text{phys}_i}),\\
\{N_1, \dots, N_m, N_{\text{cls}}\} &= \text{TAGFormer}(\{n_1, \dots, n_m\}, \mathcal{E}).
\end{split}
\end{equation}

% \textbf{Auxiliary RTL and Layout encoders} support cross-stage alignment for NetTAG, providing additional functional and physical insights from RTL and layout stages. The RTL encoder is a pre-trained LLM-based text encoder, while the layout encoder is a pre-trained graph transformer encoder.
% The pre-trained RTL encoder processes the RTL code text, generating a sequence of RTL embeddings $\{R_{\text{cls}}, R_1, \dots \}$, where $R_{\text{cls}}$ represents the embedding of the entire RTL code. 
% Similarly, the pre-trained layout encoder takes the layout physical graph as inputs, converting it into a sequence of layout embeddings $\{L_{\text{cls}}, L_1, \dots \}$, where $L_{\text{cls}}$ represents the entire layout embedding.\looseness=-1 

\textbf{Auxiliary RTL and Layout Encoders} support cross-stage alignment for NetTAG by providing functional and physical insights from RTL and layout stages. The pre-trained RTL encoder, an LLM-based text encoder, generates RTL embeddings $R_{\text{cls}}$. Similarly, the layout encoder, a pre-trained graph transformer, produces layout embeddings $L_{\text{cls}}$. Please note that these auxiliary cross-stage encoders are used only during pre-training.

\subsection{Pre-Training Step 1: Enhacing logic understanding in ExprLLM} \label{sec:pretrain1}
We design pre-training objectives tailored for circuit netlists to capture both functional and structural information using our multimodal NetTAG. 
As shown in~\Cref{fig:pretrain}, the pre-training process is divided into two main steps: (1) Pre-training ExprLLM to strengthen its understanding of Boolean logic, and (2) Pre-training TAGFormer to fuse gate text semantics from frozen ExprLLM with circuit graph structure. Additionally, two pre-trained RTL and layout encoders are used for cross-stage alignment. We detail our pre-training objectives below.\looseness=-1

% \subsubsection{Pre-training of ExprLLM}

\textbf{Objective \#1 Symbolic expression contrastive learning.} In step1, we apply logic expression contrastive learning to enhance ExprLLM’s understanding of Boolean expressions. 
As shown in~\Cref{fig:align}, we begin by building a gate expression dataset from 2-hop symbolic expressions\footnote{We choose 2-hop to balance the expression expansion and runtime.}. Then each expression is transformed using randomly applied Boolean equivalence rules\footnote{We implemented Boolean transformation rules such as De-Morgan's law, distributive law, commutative law, associative law, etc.} to generate a new positive sample. The InfoNCE loss~\cite{oord2018representation} is then applied to differentiate each positive expression pair from other negatives, formulated as:
\vspace{-.05in}
\begin{equation}\label{f:nce}
\mathcal{L}^{\#1}_{\text{expr}} = \mathcal{F}_\text{CL}(\boldsymbol{T}, T^+) = - \log \frac{\exp(T_{\text{ori}} \cdot {T_{\text{}}}^+ / \tau)}{\sum_{i=0}^{k} \exp(T_{\text{ori}} \cdot {T_i} / \tau)},
\end{equation}
where $\tau$ is the temperature scaling factor, $T^+$ is the positive embeddings, and $\boldsymbol{T}$ represents all $k$ samples in the batch, including the original sample embeddings $T_{\text{ori}}$ along with all other $(k-1)$ negative sample embeddings $T_i (i\neq \text{ori})$.

\subsection{Pre-Training Step 2: Fusion in TAGFormer \& Cross-Stage Align} \label{sec:pretrain2}

\subsubsection{Pre-training within TAGFormer for semantic and structure fusion}
We propose gate-level and graph-level self-supervised objectives to jointly pre-train TAGFormer, fusing semantic-rich gate embeddings from pre-trained and frozen ExprLLM with the global circuit structure. 
The pre-training objectives are illustrated below:\looseness=-1

\textbf{Objective \#2.1 Masked gate reconstruction.} This objective aims to capture the structural roles of various gates by leveraging their connectivity within the netlist graph. Specifically, we propose to randomly mask a subset of gates in the netlist using a special \texttt{[MASK]} node. The model is then tasked with predicting the gate type (e.g., NOR, MUX, AND, etc.) of the masked gates based on the remaining unmasked nodes.\looseness=-1 

This objective is formulated as a multi-class classification problem, where the predicted gate types are treated as discrete labels. NetTAG first processes the masked graph to generate inputs and generates masked node embeddings $N_i^{\text{mask}}$ for each masked gate in $\mathcal{T}^{\text{mask}}$. Then a classification Multi-layer Perceptron, denoted as $\text{MLP}_{\text{class}}$, 
takes the masked node embeddings $N_i^{\text{mask}}$ as input and outputs the predicted probability distribution over gate types. The ground-truth gate type for each masked node $i$ is denoted by $\boldsymbol{y}_i^{\text{mask}}$.
The classification loss is defined using the cross-entropy loss as follows:
\vspace{-.05in}
\begin{equation}
\mathcal{L}^{\#2}_{\text{gate}} = - \sum_{i \in \mathcal{T}^{\text{mask}}} \boldsymbol{y}_{i}^{\text{mask}} \log\left(\text{MLP}_{\text{class}}(N_i^{\text{mask}})\right).
\end{equation}

% \textbf{Objective \#2.2 Netlist graph contrastive learning.} This objective is designed to capture the global structural and functional information of the entire netlist graph. Specifically, we apply graph-level contrastive learning to cluster functionally similar netlists together, while pushing apart those with dissimilar functionalities.
% To augment the netlist dataset, we generate positive samples by applying functionally equivalent transformations to each netlist graph, similar to the process illustrated in~\cite{wang2024fgnn2}. Negative samples are created by directly using different graphs from the same batch, representing distinct circuit functionalities.\looseness=-1

\textbf{Objective \#2.2 Netlist graph contrastive learning.} This objective captures global structural and functional information by clustering similar netlists and separating dissimilar ones. Positive samples are generated via functionally equivalent transformations of each netlist graph~\cite{wang2024fgnn2}, while negative samples are distinct graphs within the same batch. The InfoNCE loss is also applied to minimize the distance between positive pairs while maximizing separation from negative samples:\looseness=-1 
\vspace{-.05in}
\begin{equation}
\mathcal{L}^{\#2}_{\text{graph}} = \mathcal{F}_\text{CL}(\boldsymbol{N}_{\text{cls}}, N_{\text{cls}}^+),
\end{equation}
where $N_{\text{cls}}^+$ denotes the positive sample embeddings, and $\boldsymbol{N}_{\text{cls}}$ includes the original and all other negative samples in the batch.

\textbf{Objective \#2.3 Netlist graph size prediction.}
This objective aims to predict the number of each type of gate in a given netlist graph based on the graph-level embedding, which is formulated as a regression problem.
Denote $\boldsymbol{y}^{size}$ as the ground-truth gate counts for a netlist graph $\mathcal{G}_\mathcal{N}$. We employ an auxiliary model $\text{MLP}_{\text{regr}}$, which takes the graph-level embedding $N_\text{cls}$ as input and outputs the predicted gate counts. The objective is to minimize the Mean Squared Error (MSE) between the predicted and actual gate counts, as formulated below:
\vspace{-.05in}
\begin{equation}
\mathcal{L}^{\#2}_{\text{size}} = \frac{1}{n} \sum_{i=1}^{n} \left( \boldsymbol{y}^{size}_i - \text{MLP}_{\text{regr}}(N_{\text{cls}}) \right)^2,
\end{equation}
where $n$ is the number of gate types, and each element in $\boldsymbol{y}^{size}$ corresponds to the predicted count for a specific gate type.

% \subsubsection{RTL-netlist-layout alignment}
\subsubsection{Pre-training beyond TAGFormer to enhance functional and physical awareness}
\textbf{Objective \#3 Cross-stage contrastive alignment.}
Beyond the pre-training objectives within NetTAG’s components, we introduce cross-stage alignment to integrate information from RTL-stage functionality and layout-stage physical implementation into NetTAG. 
This alignment enhances NetTAG’s cross-stage awareness, benefiting both functional and physical downstream tasks, as further demonstrated in~\Cref{sec:abl}.
Using contrastive objectives, we align the circuit embeddings from RTL (i.e., $R_{\text{cls}}^+$), netlist (i.e., $\boldsymbol{N}_{\text{cls}}$), and layout (i.e., $L_{\text{cls}}^+$) within a shared latent space, as formulated below:\looseness=-1
\vspace{-.05in}
\begin{equation}
\mathcal{L}^{\#3}_{\text{align}} = \mathcal{F}_\text{CL}(\boldsymbol{N}_{\text{cls}}, R_{\text{cls}}^+) + \mathcal{F}_\text{CL}(\boldsymbol{N}_{\text{cls}}, L_{\text{cls}}^+).
\end{equation}

To this end, we formulate the overall self-supervised pre-training objective of NetTAG as the following two-step process: 
\vspace{-.05in}
\begin{equation}
\mathcal{L}_{\text{NetTAG}} =
\begin{cases}
\mathcal{L}^1_{\text{expr}} & \text{(Step 1)} \\
\mathcal{L}^2_{\text{gate}} + \mathcal{L}^2_{\text{graph}} + \mathcal{L}^2_{\text{size}} + \mathcal{L}^3_{\text{align}} & \text{(Step 2)}
\end{cases}
\end{equation}

\subsection{Fine-tuning for downstream tasks} \label{sec:finetune}
NetTAG generates multi-grained embeddings for netlist elements, including combinational gates, register cones, and entire circuits. For sequential circuits, circuit-level embeddings are computed by summing the embeddings of all register cones, while for combinational circuits, their overall embeddings are directly obtained on the \texttt{[CLS]} node without the need for chunking into cones.

For downstream tasks, we fine-tune these embeddings with lightweight task models like MLPs or tree-based models (e.g., XGBoost). NetTAG’s versatile embeddings support both regression and classification across functional and physical tasks for netlists. For functional tasks, the fine-tuning model predicts early RTL-stage information on netlists, such as gate functions and register types. As for physical tasks, it predicts late layout-stage metrics like register endpoint slack, power, and area. We will provide the detailed evaluation results in~\Cref{sec:tasks}.

\section{Experimental Results} \label{sec:expr}

\subsection{Experimental Setup}
   
\textbf{Data preparation.}
We collect circuits for pre-training from various sources, including benchmark RTL code from ITC99~\cite{corno2000rt}, OpenCores~\cite{URL:opencore}, Chipyard~\cite{amid2020chipyard}, and VexRiscv~\cite{vexriscv}. All RTL designs are synthesized into netlists using Synopsys Design Compiler using NanGate 45nm technology library and then undergo physical design with Cadence Innovus. Gate and design quality metrics and statistics are obtained using Synopsys PrimeTime.
For symbolic logic expressions, we use PySMT~\cite{gario2015pysmt}, a symbolic reasoning toolkit for formal verification, to construct and manipulate the Boolean expressions.

The statistical details of our dataset are shown in~\Cref{tbl:stat}. For the expression dataset used by ExprLLM, we collected 313k original expressions, each augmented with functionally equivalent transformations, resulting in a total of 626k expressions. As for the circuit netlist data, we collected 100k subcircuit cones after chunking, which were also functionally augmented to reach 200k samples in total. Additionally, we included 10k aligned RTL and layout cones for cross-stage alignment. When applying NetTAG to downstream tasks, we directly use the original task datasets if available. Otherwise, we utilize designs from the aforementioned open-source benchmarks.

\begin{table}[!h]
\centering
\vspace{-.05in}
\caption{Statistics of circuit expression and netlist dataset.}
\vspace{-.05in}
\resizebox{0.4\textwidth}{!}{

\begin{tabular}{c|cc|cc} \toprule
\multirow{3}{*}{\textbf{Source}} & \multicolumn{2}{c|}{\textbf{Gate Expression}}                                     & \multicolumn{2}{c}{\textbf{Circuit Netlist}}                             \\
                                                                                              & \# Data & \begin{tabular}[c]{@{}c@{}}\# Tokens \\      \scriptsize(Avg.)\end{tabular} & \# Data & \begin{tabular}[c]{@{}c@{}}\# Nodes\\      \scriptsize(Avg.)\end{tabular} \\ \hline \hline
ITC99                                                                                         & 47k        & 6,960                                                            & 4k   & 1,025                                                          \\
OpenCores                                                                                     & 76k        & 212                                                              & 55k  & 173                                                            \\
Chipyard                                                                                      & 109k       & 9,849                                                            & 20k  & 2,813                                                          \\
VexRiscv                                                                                      & 81k        & 5,289                                                            & 21k  & 901                                                            \\ \hline
\textbf{Total}                                                                                & \textbf{313k}       & \textbf{5,810}                                                            & \textbf{100k}  & \textbf{855} \\ \bottomrule                                                          
\end{tabular}

}

\label{tbl:stat}
\vspace{-.1in}
\end{table}

\textbf{Implementation details.}
For the implementation of NetTAG, we initialize ExprLLM with LLM2Vec~\cite{behnamghader2024llm2vec}, an open-source LLM-based encoder based on Meta Llama-3.1-8B~\cite{dubey2024llama}, with 8k maximum input tokens. For the graph transformer backbone of TAGFormer, we adopt SGFormer~\cite{wu2024simplifying}. Additionally, we use NV-Embed~\cite{lee2024nv} (maximum 32k input tokens) as the pre-trained RTL text encoder, and the layout encoder is pre-trained using a graph contrastive objective on SGFormer. Each MLP contains three layers with a hidden dimension of 256, and the output dimension of NetTAG is set to 768.
We pre-train ExprLLM using LoRA on the full dataset for one epoch, followed by 50 epochs of pre-training for TAGFormer. Afterward, fine-tuning is performed with lightweight models and task-specific labels. All experiments are conducted on 8 Nvidia 4090 GPUs and 4 Nvidia A6000 GPUs.\looseness=-1 

\subsection{Performance on Various Downstream Tasks}\label{sec:tasks}
We evaluate the pre-trained NetTAG on four representative netlist-stage tasks, including functional and physical tasks across combinational and sequential circuits. These tasks span multiple netlist granularities: combinational gates, register cones, and entire circuits, offering a thorough assessment of NetTAG’s capabilities.\looseness=-1 

NetTAG consistently outperforms task-specific baselines in each task, demonstrating its ability to capture functional and physical circuit information. This highlights NetTAG's role as a foundation model that is capable of generating informative embeddings for netlists.
Detailed evaluations for each task are as follows:

\textbf{Task1: Combinational gate function identification.} This task identifies the functional type of each netlist combinational gate (e.g., adder, multiplier, comparator) as described in the original RTL code. Accurate gate function identification is critical for applications such as reverse engineering, hardware security, and functional verification. We evaluate NetTAG on an open-source dataset from GNN-RE~\cite{alrahis2021gnn}, ensuring no label-related text is included in the gate text attributes and using the same metrics for a fair comparison.

As shown in~\Cref{tbl:task1}, NetTAG significantly outperforms GNN-RE across all metrics, with an average improvement of 14\% in accuracy (97\% vs. 83\%), 11\% in precision, 14\% in recall and F1-score. These improvements demonstrate NetTAG’s robust ability to capture functional information on each gate, enabling superior generalization compared to the task-specific GNN-based method.

\begin{table}[!h]
\centering
\caption{Performance comparison on Task1: combinational gate function identification.}
\vspace{-.05in}
\resizebox{0.45\textwidth}{!}{

\begin{tabular}{c||cccc||cccc} \toprule
       & \multicolumn{4}{c||}{\cellcolor[HTML]{F2F2F2}\textbf{GNN-RE~\cite{alrahis2021gnn}}} & \multicolumn{4}{c}{\cellcolor[HTML]{FFF2CC}\textbf{NetTAG}}                       \\ 
 & Acc.   & Prec.   & Recall  & F1  & Acc.      & Prec.     & Recall        & F1      \\ 
\multirow{-3}{*}{Design} & \scriptsize(\%) & \scriptsize(\%) & \scriptsize(\%) & \scriptsize(\%) & \scriptsize(\%) & \scriptsize(\%) & \scriptsize(\%) & \scriptsize(\%) \\
\hline \hline
1      & 79& 82& 79& 74& 97& 97& 97& 97\\
2      & 96& 96& 96& 96& 100& 100& 100& 100\\
3      & 94& 94& 94& 94& 100& 100& 100& 100\\
4      & 78& 83& 78& 78& 100& 100& 100& 100\\
5      & 91& 92& 91& 90& 99& 99& 99& 99\\
6      & 74& 78& 74& 68& 94& 94& 94& 93\\
7      & 80& 80& 80& 80& 84& 87& 84& 81\\
8      & 89& 90& 89& 87& 95& 96& 95& 96\\
9      & 65& 77& 65& 67& 100& 100& 100& 100\\ \hline
\textbf{Avg.}    & \cellcolor[HTML]{F2F2F2}83& \cellcolor[HTML]{F2F2F2}86        & \cellcolor[HTML]{F2F2F2}83    & \cellcolor[HTML]{F2F2F2}82      & \cellcolor[HTML]{FFF2CC}\textbf{97}& \cellcolor[HTML]{FFF2CC}\textbf{97}& \cellcolor[HTML]{FFF2CC}\textbf{97}& \cellcolor[HTML]{FFF2CC}\textbf{96}\\ \bottomrule
\end{tabular}

}

\label{tbl:task1}
\vspace{-.1in}
\end{table}

\textbf{Task2: Sequential state/data register identification.}
This task distinguishes state registers from data path registers, which is essential for understanding high-level control and data flow within low-level netlists, supporting security analysis and verification.
We evaluate NetTAG against ReIGNN~\cite{chowdhury2021reignn}, a supervised method specifically designed for this task based on GNN. The evaluation metrics include sensitivity and balanced accuracy (the average of sensitivity and true negative rate).\looseness=-1

As demonstrated in~\Cref{tbl:task2}, NetTAG significantly outperforms ReIGNN in both metrics. It achieves a 44\% improvement in sensitivity (90\% vs. 46\%) for state register identification and a 13\% increase in overall accuracy (86\% vs. 73\%), highlighting its superior capability of capturing structural and functional information from register cones.\looseness=-1

\begin{table}[!h]
\centering
\vspace{-.1in}
\caption{Performance comparison on Task2: state/data register identification \& Task3: endpoint register slack prediction.}
\vspace{-.05in}

\resizebox{0.5\textwidth}{!}{
\begin{tabular}{c||cc||cc||cc||cc} \toprule
\multicolumn{1}{c||}{}     & \multicolumn{4}{c||}{\textbf{Task 2}}         & \multicolumn{4}{c}{\textbf{Task 3}}   \\ \cline{2-9}
 &
  \multicolumn{2}{c||}{\cellcolor[HTML]{F2F2F2}\textbf{REIGNN~\cite{chowdhury2021reignn}}} &
  \multicolumn{2}{c||}{\cellcolor[HTML]{FFF2CC}\textbf{NetTAG}} &
  \multicolumn{2}{c||}{\cellcolor[HTML]{F2F2F2}\textbf{GNN$^\star$}} &
  \multicolumn{2}{c}{\cellcolor[HTML]{FFF2CC}\textbf{NetTAG}} \\ 
 &
  Sens. &
  Acc. &
  Sens. &
  Acc. &
  \multicolumn{1}{c}{R} &
  \multicolumn{1}{c||}{MAPE} &
  R &
  MAPE \\ 
\multirow{-4}{*}{Design} & \scriptsize(\%) & \scriptsize(\%) & \scriptsize(\%) & \scriptsize(\%) &  & \scriptsize(\%) &  & \scriptsize(\%) \\  
  \hline \hline
itc1       & 50  & 72 & 100            & 98          &  0.89&  13& 0.94& 9\\
itc2       & 100    & 92 & 100            & 100             &  0.91&  10& 0.93& 9\\
chipyard1  & 30  & 65 & 80          & 79          &  0.72&  17& 0.7& 20\\
chipyard2  & 30  & 65 & 90          & 86          &  0.86&  12& 0.95& 9\\
vex1       & 50  & 74 & 82         & 74          &  0.94&  11& 0.93& 12\\
vex2       & 32 & 60  & 86         & 82          &  0.97&  18& 0.97& 9\\
opencores1 & 42 & 73 & 93         & 84          &  0.99&  26& 0.92& 29\\
opencores2 & 37 & 80  & 92         & 82          &  0.93&  30& 0.98& 26\\ \hline
\textbf{Avg.}        & \cellcolor[HTML]{F2F2F2}46 & \cellcolor[HTML]{F2F2F2}73 & \cellcolor[HTML]{FFF2CC}\textbf{90} & \cellcolor[HTML]{FFF2CC}\textbf{86} &  0.9&  17& \cellcolor[HTML]{FFF2CC}\textbf{0.92}& \cellcolor[HTML]{FFF2CC}\textbf{15}\\ \bottomrule
\end{tabular}

}

\begin{tablenotes}\footnotesize
\item $^\star$ We adapt the GNN model from~\cite{wang2023restructure} for netlist-stage slack prediction, as it was originally designed for the layout stage.
\end{tablenotes} 

\label{tbl:task2}
\vspace{-.2in}
\end{table}

\textbf{Task3: Endpoint register slack prediction.}
This task focuses on predicting sign-off timing slack at the netlist stage to provide early feedback, thereby expanding the optimization space in the physical design flow. The prediction is highly challenging due to substantial graph topology changes during physical design optimizations, as discussed in~\cite{wang2023restructure}.
We compare NetTAG with the state-of-the-art GNN-based method from~\cite{wang2023restructure}, extending their post-placement method to work at the post-synthesis stage. Performance is evaluated using correlation coefficient (R) and mean absolute percentage error (MAPE).\looseness=-1

As shown in~\Cref{tbl:task2}, NetTAG outperforms the customized GNN baseline~\cite{wang2023restructure}, achieving more accurate slack prediction with R of 0.92 vs. 0.9 and a MAPE of 15\% vs. 17\%.

\textbf{Task4: Overall circuit power/area prediction.} 
This task predicts final layout power and area metrics at the netlist stage, offering early estimates for physical design. We compare NetTAG with the synthesis tool and a customized GNN model adopted from PowPrediCT~\cite{du2024powpredict}, a layout-stage state-of-the-art optimization-aware power predictor.

For area prediction, NetTAG achieves a 4\% MAPE without optimization and 11\% with optimization, outperforming the GNN’s 5\% and 18\% and the synthesis EDA tool’s 5\% and 34\%. For power, it yields an 8\% MAPE without optimization and 12\% with optimization, surpassing the GNN’s 12\% and 19\% and the EDA tool’s 34\% and 38\%. These results highlight NetTAG’s strong accuracy and robustness for circuit-level tasks.\looseness=-1

\begin{table}[!h]
\centering
\caption{Performance comparison on Task4: overall circuit power/area prediction.}
\vspace{-.05in}
\resizebox{0.45\textwidth}{!}{

\begin{tabular}{cc||cc||cc||cc} \toprule 
\multicolumn{2}{c||}{\multirow{3}{*}{Target Metric$^\dagger$}} &
  \multicolumn{2}{c||}{\cellcolor[HTML]{F2F2F2}\textbf{EDA Tool}} &
  \multicolumn{2}{c||}{\cellcolor[HTML]{F2F2F2}\textbf{GNN$^\star$}} &
  \multicolumn{2}{c}{\cellcolor[HTML]{FFF2CC}\textbf{NetTAG}} \\
\multicolumn{2}{c||}{}             & R    & MAPE & R & MAPE & R    & MAPE \\ 
& & & \scriptsize(\%) & & \scriptsize(\%) & & \scriptsize(\%)
\\ \hline \hline
\multirow{2}{*}{Area}  & w/o opt & 0.99 & 5    &   0.99&      5& \textbf{0.99} & \textbf{4}    \\
                       & w/ opt  & 0.95 & 34   &   0.95&      18& \textbf{0.96} & \textbf{11}   \\ \hline
\multirow{2}{*}{Power} & w/o opt & 0.99 & 34   &   0.99&      12& \textbf{0.99} & \textbf{8}    \\ 
                       & w/ opt  & 0.73 & 38   &   0.76&      19& \textbf{0.86} & \textbf{12}  \\ \bottomrule
\end{tabular}

}
\begin{tablenotes}\footnotesize
\item $^\dagger$ Post-layout labels are collected in two scenarios: without considering optimization (denoted as ``w/o opt'') and with optimization as illustrated in~\cite{du2024powpredict} (denoted as ``w/ opt'').
\item $^\star$ We adapt the GNN model from~\cite{du2024powpredict} for netlist power and area prediction, as it was originally designed for the layout-stage power prediction.
\end{tablenotes} 

\label{tbl:task4}
\vspace{-.2in}
\end{table}

% \textbf{Performance summary.} 
% The pre-trained NetTAG demonstrates strong adaptability across a range of functional and physical tasks, spanning multiple circuit granularities and gate types. In every task evaluation, NetTAG significantly outperforms task-specific baselines, underscoring its capacity to capture both structural and semantic information essential for netlist-stage tasks. This performance advantage highlights NetTAG’s versatility and potential as a foundation netlist model for complex circuit analysis.

\begin{figure}[!b]
  \centering
  \includegraphics[width=1\linewidth]{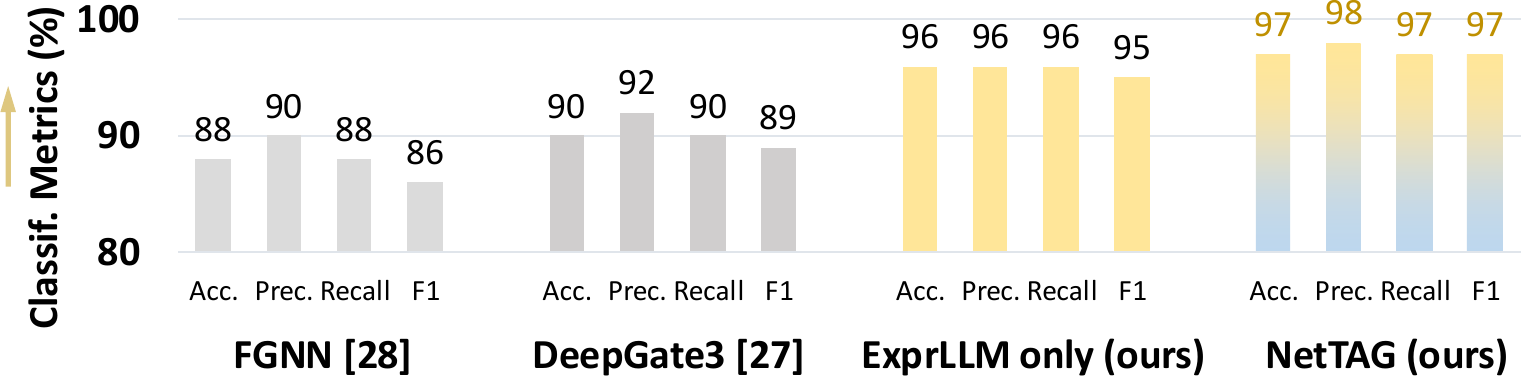}
  \vspace{-.2in}
  \caption{Comparision with pre-trained AIG encoders on AIG dataset.}
  \label{fig:encoder}
  \vspace{-.2in}
\end{figure}

\subsection{Comparision with pre-trained netlist encoders}
In addition to the task-specific methods, we also compare NetTAG with pre-trained netlist encoders. Since they are all limited to combinational AIG circuits, we evaluate them only on Task 1, using an AIG-format dataset.
As shown in~\Cref{fig:encoder}, NetTAG outperforms SOTA AIG-based encoders, including FGNN~\cite{wang2024fgnn2} and DeepGate3~\cite{shi2024deepgate3}, achieving the highest average performance across all metrics. Notably, NetTAG achieves superior accuracy on the AIG dataset compared with more diverse gate datasets in~\Cref{tbl:task1} for Task 1, underscoring its adaptability.
We also evaluate the standalone ExprLLM component of NetTAG, which performs well using symbolic expressions, highlighting the power of gate semantic understanding. These results emphasize NetTAG’s advantage in combining semantic and structural information through the TAG format.

% As shown in~\Cref{fig:encoder}, NetTAG is compared with the leading AIG-based encoders, including FGNN~\cite{wang2024fgnn2}, a GNN-based method, and DeepGate3~\cite{shi2024deepgate3}, which leverages a graph transformer. NetTAG achieves the highest average performance across all metrics, significantly outperforming both FGNN and DeepGate3. Notably, NetTAG achieves superior accuracy on the AIG dataset for Task 1, compared to results on more diverse gate datasets in~\Cref{tbl:task1}, demonstrating its robustness and adaptability in AIG format.

% To further investigate, we evaluate the pre-trained ExprLLM component of NetTAG independently. Remarkably, ExprLLM alone, based solely on textual symbolic expression, achieves strong performance, underlining the impact of semantic understanding on circuit functionality. The results highlight NetTAG’s advantage in integrating semantic and structural information via TAG format.\looseness=-1 

\subsection{Ablation Study}\label{sec:abl}
\begin{figure}[!t]
  \centering
  \vspace{-.1in}
  \includegraphics[width=0.86\linewidth]{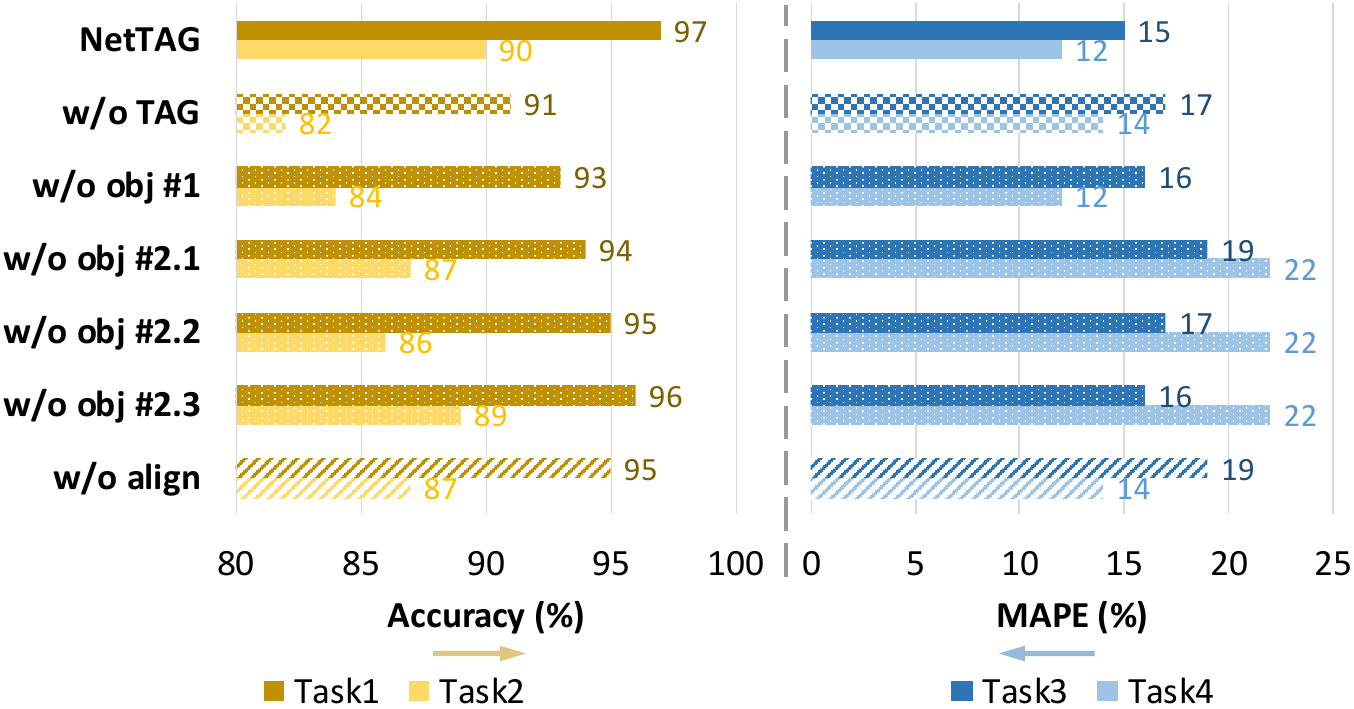}
  \vspace{-.05in}
  \caption{Ablation study.}
  \label{fig:abl}
  \vspace{-.2in}
\end{figure}

This ablation study evaluates the contributions of three key aspects of NetTAG: TAG-based representation, self-supervised pre-training objectives, and cross-stage alignment. The analysis, shown in~\Cref{fig:abl}, highlights the impact of each component on the model’s performance across the four tasks, with detailed analysis as follows:

\textbf{Effectiveness of learning netlist via TAG.} Removing text attributes and relying only on graph structure significantly reduces performance, especially on functional tasks. This indicates that semantic information from text attributes is crucial. Additionally, physical tasks also show a slight decline, suggesting that even structure-focused tasks benefit from the added semantic context in TAG.

\textbf{Effectiveness of self-supervised pre-training objectives.} 
The results demonstrate that the expression contrastive learning objective (\#1) for ExprLLM has the greatest impact on functional tasks, suggesting it enhances the LLM’s understanding of symbolic logic expressions. Objectives \#2.1 (masked gate reconstruction) and \#2.2 (graph contrastive learning) improve performance across both functional and physical tasks, highlighting their role in capturing both local and global netlist structure. Objective \#2.3 (graph size prediction) shows the strongest effect on physical tasks, where a comprehensive understanding of the netlist’s overall structure is essential.

\textbf{Effectiveness of cross-stage alignment.} Removing cross-stage alignment leads to a notable drop in performance across all four tasks. This alignment effectively enhances NetTAG’s ability to integrate both early-stage functional and late-stage physical information, which is crucial for downstream tasks.

\begin{table}[!b]
\centering
\caption{Runtime (minutes) Comparision.}
\vspace{-.05in}
\resizebox{0.45\textwidth}{!}{
\begin{tabular}{c||c||c|c|c|c} \toprule
\multirow{3}{*}{Source} & \textbf{EDA Tool}                                                                      & \multicolumn{4}{c}{\textbf{Ours (Avg.)}}                                                  \\ \cline{2-6}
                        & \multirow{2}{*}{\begin{tabular}[c]{@{}c@{}}P\&R   \\      (Avg.)\end{tabular}} & \multirow{2}{*}{Pre$^\star$} & \multicolumn{2}{c|}{Infer} & \multirow{2}{*}{Total} \\ \cline{4-5}
                        &                                                                               &                      & ExprLLM    & TAGFormer    &                        \\ \hline \hline
ITC99                   & 164                                                                           & 2                    & 5          & 0            & 7                      \\
OpenCores               & 288                                                                           & 18                   & 12         & 1            & 31                     \\
Chipyard                & 251                                                                           & 15                   & 10         & 1            & 26                     \\
VexRiscv                & 207                                                                           & 8                    & 5          & 2            & 15                     \\
GNNRE                   & /                                                                             & 4                    & 2          & 0            & 6         \\ \bottomrule            
\end{tabular}

}
\begin{tablenotes}\footnotesize
\item $^\star$ Preprocessing (chunking into cones and converting netlist into TAG).
\end{tablenotes} 

\label{tbl:runtime}
\vspace{-.2in}
\end{table}

\subsection{Runtime Analysis}
In~\Cref{tbl:runtime}, we present NetTAG’s runtime across various benchmarks. Most runtime is spent on preprocessing (i.e., converting netlists into TAG format) and node-level inference using ExprLLM. Despite these steps, NetTAG demonstrates around a 10x speedup over traditional physical design workflows using commercial EDA tools.
There is significant potential for further runtime improvements. For example, symbolic expression extraction for each gate can be highly parallelized, reducing preprocessing time. Additionally, the ExprLLM inference time could be largely reduced by scaling up GPU resources, as this runtime depends on GPU performance and quantity.
 
\section{Discussion}

\subsection{Scalability: Performance Scaling with Model and Data Size}
\begin{figure}[!h]
  \centering
  \includegraphics[width=0.9\linewidth]{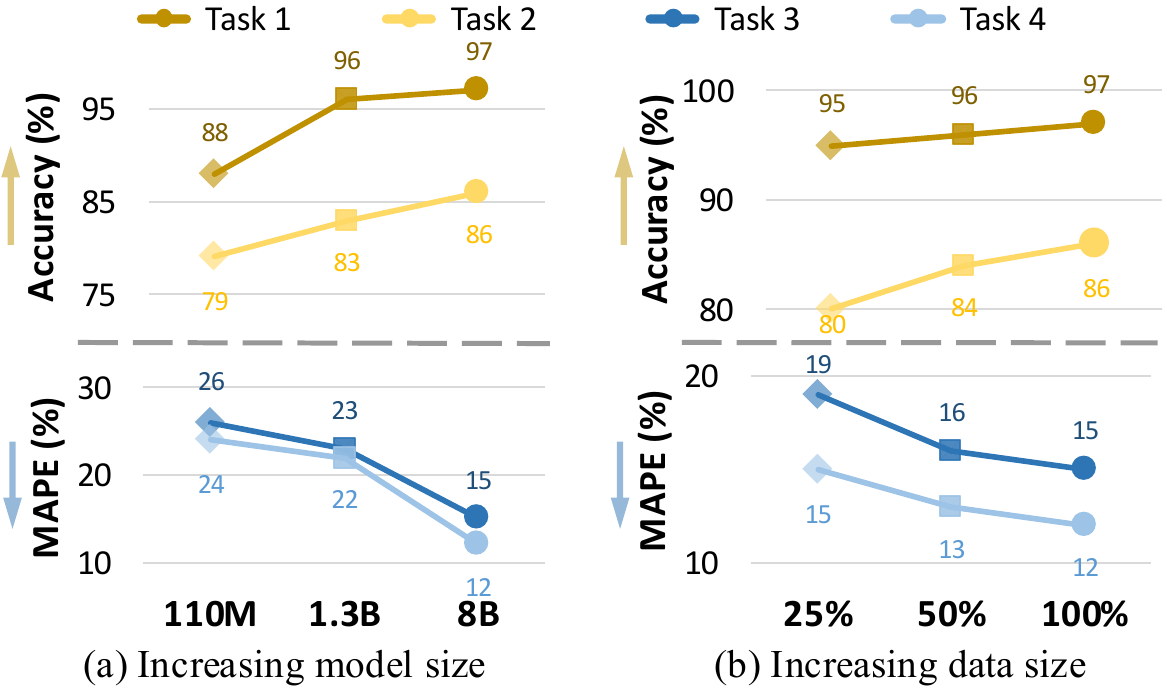}
  \vspace{-.05in}
  \caption{Performance scaling with model and data size.}
  \label{fig:scale}
  \vspace{-.1in}
\end{figure}

In~\Cref{fig:scale}, we study how the downstream task performance of NetTAG scales with both model size and pre-training data size. The plot shows the performance of each task after fine-tuning.
Scaling ExprLLM backbone from 110M parameters (BERT) to larger models like Meta Llama 3.1 with 1.3B and 8B parameters results in significant performance improvements across all four tasks. Similarly, expanding data size from 25\% to 100\% of the dataset consistently enhances performance. These results demonstrate the scalability of NetTAG, suggesting that further increases in model and data size could lead to even greater improvements in accuracy and generalization.\looseness=-1

% Specifically, when the model size is increased from 110M to 8B parameters\footnote{We scale the LLM-based encoder from BERT (110M) to Llama 3.1-1.5B and Llama 3.1-8B}, accuracy improves from 79\% to 96\% on Task 1 and from 83\% to 97\% on Task 2. Physical tasks (Tasks 3 and 4) also show improvement, with MAPE decreasing from 24\% to 12\% for Task 3 and from 22\% to 19\% for Task 4. This trend indicates that larger models are more capable of capturing complex circuit semantics and structural details, leading to enhanced performance.
% Similarly, scaling the data size also leads to a noticeable improvement in performance. Scaling the pre-training data size from 25\% to 100\% .., highlighting the importance of using a large, diverse circuit dataset for pre-training.

\subsection{Demo: Reasoning Netlist's Arithmetic Function with NetTAG}
\begin{figure}[!h]
  \centering
  \includegraphics[width=1\linewidth]{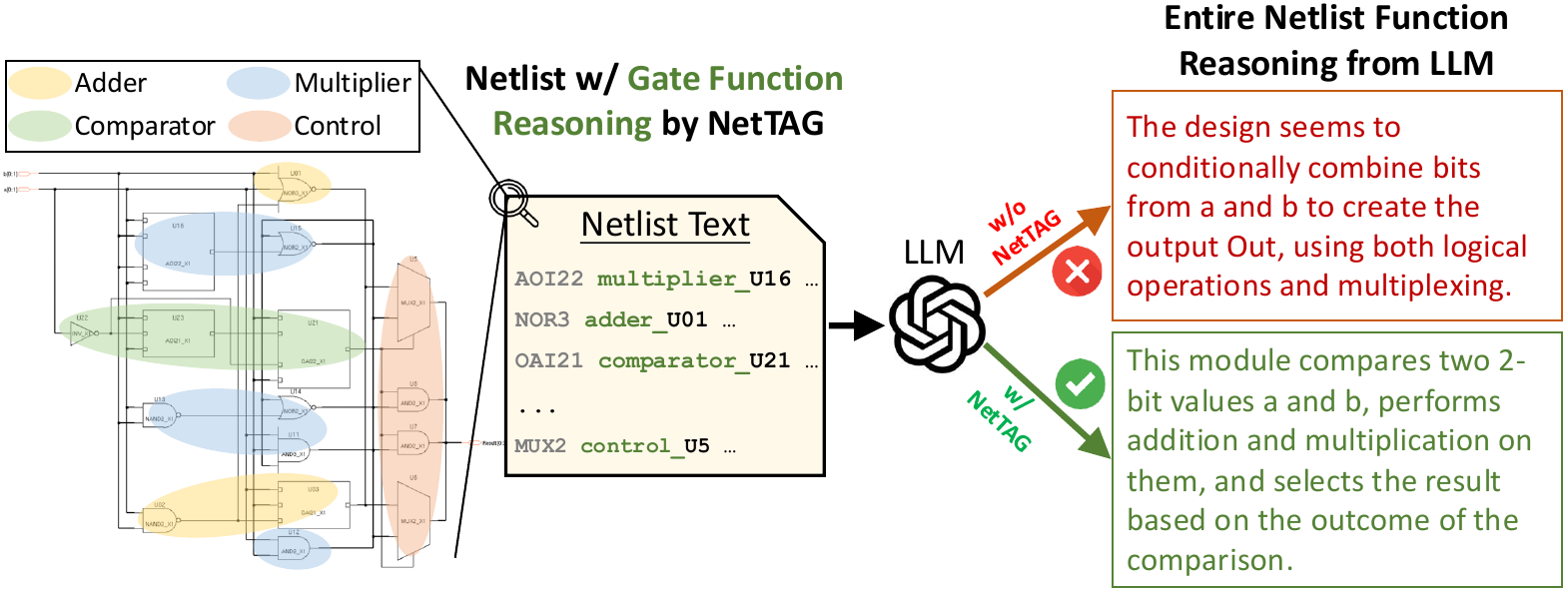}
  \vspace{-.2in}
  \caption{A demo for reasoning arithmetic function from netlists.}
  \label{fig:demo1}
  \vspace{-.1in}
\end{figure}

\Cref{fig:demo1} shows how NetTAG improves functional reasoning for arithmetic netlists. Using an LLM (e.g., OpenAI’s o1-preview) to analyze post-synthesis netlist Verilog texts, we assess its ability to interpret the arithmetic function from the flattened netlists.
Without NetTAG’s gate-level function identification, the LLM struggles with interpreting functionality from flattened netlists. When integrated with NetTAG, the LLM accurately infers that the netlist module compares two 2-bit values, performs addition and multiplication, and selects the result based on the comparison outcome.

\section{Conclusion And Future Work}
In this paper, we present NetTAG, a foundation model for netlist representation learning that accommodates diverse gate types and supports a range of functional and physical tasks. By formulating netlists as text-attributed graphs, NetTAG uniquely integrates gate semantics with graph structure through its multimodal architecture and self-supervised pre-training objectives. Future work will enhance NetTAG by extending pre-training to larger circuit datasets and exploring decoding methods to broaden task support capabilities.\looseness=-1

\section{Acknowledgement}
This work is supported by Hong Kong Research Grants Council (RGC) CRF Grant C6003-24Y, National Natural Science Foundation of China (NSFC) 62304192, and ACCESS – AI Chip Center for Emerging Smart Systems, sponsored by InnoHK, Hong Kong SAR. 
We thank HKUST Fok Ying Tung Research Institute and National Supercomputing Center in Guangzhou Nansha Sub-center for computational resources.

%The authors would like to thank HKUST Fok Ying Tung Research Institute and National Supercomputing Center in Guangzhou Nansha Sub-center for providing high-performance computational resources. 

\clearpage
\bibliographystyle{IEEEtran}
\bibliography{ref}

\end{document}